\documentclass[prb,twocolumn,showkeys,showpacs,preprintnumbers,amsmath,amssymb,floatfix]{revtex4}

\usepackage{graphicx}
\usepackage{dcolumn}
\usepackage{bm}

\begin{document}

\title{Advanced resistivity model for arbitrary magnetization orientation applied to a series
of compressive- to tensile-strained (Ga,Mn)As layers}

\author{W. Limmer}
\email{wolfgang.limmer@uni-ulm.de}
 \homepage{http://hlpsrv.physik.uni-ulm.de}
\author{J. Daeubler}
\author{L. Dreher}
\author{M. Glunk}
\author{W. Schoch}
\author{S. Schwaiger}
\author{R. Sauer}
\affiliation{Institut f\"ur Halbleiterphysik, Universit\"at Ulm, 89069 Ulm, Germany}

\date{\today}

\begin{abstract}
The longitudinal and transverse resistivities of differently strained (Ga,Mn)As layers are
theoretically and experimentally studied as a function of the magnetization orientation.
The strain in the series of (Ga,Mn)As layers is gradually varied from compressive to tensile
using (In,Ga)As templates with different In concentrations. Analytical expressions for the
resistivities are derived from a series expansion of the resistivity tensor with respect to
the direction cosines of the magnetization. In order to quantitatively model the experimental
data, terms up to the fourth order have to be included. The expressions derived are generally
valid for any single-crystalline cubic and tetragonal ferromagnet and apply to arbitrary surface
orientations and current directions. The model phenomenologically incorporates the longitudinal
and transverse anisotropic magnetoresistance as well as the anomalous Hall effect. The resistivity
parameters obtained from a comparison between experiment and theory are found to systematically
vary with the strain in the layer.
\end{abstract}



\maketitle

\section{Introduction}

The implementation of ferromagnetism in III-V semiconductors by incorporating high concentrations
of magnetic elements into the group-III sublattice has opened the prospect of extending
conventional semiconductor technology to magnetic applications.\cite{Ohn98,Mac05,Jun06}
A prominent example, which is presently intensely studied, is the diluted ferromagnetic
semiconductor (Ga,Mn)As. Even though the Curie temperatures reported so far are well below
room temperature, it represents a potential candidate or at least an ideal test system for
spintronic applications due to its compatibility with the standard semiconductor GaAs. During
the last decade, considerable progress has been made in understanding the basic structural,
electronic, and magnetic properties of (Ga,Mn)As.\cite{Jun06} Longitudinal anisotropic
magnetoresistance (AMR)\cite{Bax02,Jun03,Mat04,Goe05,Wan05a} and transverse AMR, often called
planar Hall effect (PHE),\cite{Tan03} anomalous Hall effect (AHE),\cite{Berg80,Jun02a,Edm03}
and magnetic anisotropy (MA)\cite{Abo01,Die01} have been established as characteristic features,
making (Ga,Mn)As potentially suitable for field-sensitive devices and non-volatile
memories.\cite{Pri98,Zut04,Pea05} Great effort has been made to understand the microscopic
mechanisms behind the observed magnetic phenomena and to obtain, theoretically and experimentally,
values for the corresponding physical parameters. In particular, the storage and processing of
information by manipulating the magnetization as well as the readout via electrical signals
demand a precise knowledge of the parameters controlling the MA and the AMR. It has been shown
that both the MA and the AMR are affected by temperature, hole density, and
strain.\cite{Ohn98,Jun02b,Liu06,Saw04,Saw05,Ham03,Ham06} For example, changing the strain from
compressive to tensile, the surface normal which usually represents a magnetic hard axis in
compressively strained (Ga,Mn)As layers turns into an easy axis in tensile-strained
layers.\cite{Ohn98,Liu06,Dae08}

As demonstrated in Ref.~\onlinecite{Lim06}, angle-dependent magnetotransport measurements,
performed at different strengths of an external magnetic field, are a genuine alternative to
ferromagnetic resonance (FMR) spectroscopy\cite{Goe03,Liu06} for probing the MA in (Ga,Mn)As.
The application of this method, however, requires analytical expressions for the longitudinal
and transverse resistivities $\rho_{\mathrm{long}}$ and $\rho_{\mathrm{trans}}$, respectively,
correctly describing the AMR and the AHE. Based on symmetry considerations, such expressions
can be derived in a phenomenological way by writing the resistivity tensor as a series expansion
with respect to the direction cosines of the magnetization. In Ref.~\onlinecite{Lim06} the
angle-dependent magnetotransport data of compressively strained (Ga,Mn)As layers, grown on
(001)- and (113)A-oriented GaAs substrates, could be well simulated considering only terms
up to the second order.

In the present work, we systematically study the influence of vertical strain on the AMR and
the AHE, investigating a series of compressive- to tensile-strained (Ga,Mn)As layers, grown on
(In,Ga)As templates with different In contents. An advanced macroscopic model is presented which
phenomenologically describes the dependence of $\rho_{\mathrm{long}}$ and $\rho_{\mathrm{trans}}$
on the magnetization orientation for cubic ferromagnets with tetragonal distortion along [001].
The analytic expressions for $\rho_{\mathrm{long}}$ and $\rho_{\mathrm{trans}}$ are first
discussed for a variety of configurations with in-plane and out-of-plane magnetization and are
then used to analyze the angle-dependent resisivities recorded from the (Ga,Mn)As layers under
study. In contrast to Ref.~\onlinecite{Lim06}, distinct features in the longitudinal out-of-plane
AMR occured which can only be described by taking into account terms up to the fourth order in
the magnetization components. Finally, the resistivity parameters, determined by fitting the
calculated curves to the experimental data, are discussed as a function of the vertical strain
in the layer. The model presented in this work applies not only to (Ga,Mn)As but most generally
to single-crystalline cubic and tetragonal ferromagnets.

\section{Experimental Details}

A series of differently strained (Ga,Mn)As layers with constant thickness of $\sim$\,180\,nm and
Mn concentration of $\sim$\,5\% was grown by low-temperature molecular-beam epitaxy on (In,Ga)As
templates in the following way: After thermal deoxidation, a 30 nm thick GaAs buffer layer was
grown at a substrate temperature of $T_\mathrm{S} \approx$ 580\,$^\circ$C on semi-insulating
GaAs(001). Then the growth was interrupted, $T_\mathrm{S}$ was lowered to $\sim$\,430\,$^\circ$C,
and a graded (In,Ga)As layer with a thickness between 0\,$\mu$m and 5\,$\mu$m was deposited
following the method described in Ref.~\onlinecite{Har89}. In order to minimize the number of
threading dislocations and to end up with different lateral lattice constants in the (In,Ga)As
templates, the In content was continuously increased in each template from 2\% up to a maximum
value of 13\%. Prior to the epitaxy of (Ga,Mn)As, the growth was again interrupted and
$T_\mathrm{S}$ was lowered to $\sim$\,250\,$^\circ$C. High-resolution x-ray diffraction (HRXRD)
reciprocal space mapping (RSM) of the (224) reflex was used to determine the vertical strain
$\varepsilon_{zz} = (a_\perp-a_\mathrm{rel})/a_\mathrm{rel}$ of the (Ga,Mn)As layers, where the
relaxed lattice constants $a_\mathrm{rel}$ were derived from the lateral and vertical lattice
constants $a_\parallel$ and $a_\perp$, respectively, applying Hooke's law. The values of
$\varepsilon_{zz}$ were found to gradually vary from $+0.24\%$ for the compressive-strained sample
without (In,Ga)As template to $-0.46\%$ for the tensile-strained sample with 13\% In. Moreover,
RSM showed that the (In,Ga)As layers were almost completely relaxed whereas the (Ga,Mn)As layers
were fully strained. Further details of the growth procedure and the RSM method will be presented
elsewhere.

For the magnetotransport studies two types of Hall bars with current directions along [100] and
[110] were prepared on several pieces of the cleaved samples. The width of the Hall bars is
0.3 mm and the longitudinal voltage probes are separated by 1 mm. High-field magnetotransport
measurements (up to 14.5\,T) at 4.2\,K yielded hole densities for the as-grown samples between
$3 \times 10^{20}$\,cm$^{-3}$ and $4 \times 10^{20}$\,cm$^{-3}$. Least squares fits were performed
to separate the contributions of the normal and anomalous Hall effect. Curie temperatures between
61\,K and 83\,K were estimated from the peak positions of the temperature-dependent sheet
resistivities at 10 mT. For the angle-dependent magnetotransport measurements, carried out at
4.2\,K, the Hall bars were mounted on the sample holder of a liquid-He-bath cryostat which was
positioned between the poles of an electromagnet system providing a maximum field strength of
0.68\,T. The sample holder has two perpendicular axes of rotation, allowing for any orientation
of the Hall bars with respect to the applied magnetic field $\bm{H}$.

\section{Theoretical model}

The macroscopic theoretical model presented in this paper is based on the assumption that the
sample area probed by magnetotransport can be approximately treated as a single homogeneous
ferromagnetic domain. It provides analytical expressions for the electrical resistivities as a
function of the magnetization orientation. Although the single-domain picture is known to usually
break down in situations where the magnetic system is undergoing a magnetization reversal process,
it has been successfully applied to the description of a variety of magnetization-related
phenomena in (Ga,Mn)As, particularly at sufficiently high external magnetic fields.

Given a single ferromagnetic domain, the macroscopic magnetization is described by the vector
$\bm{M}=M\bm{m}$ where $M$ denotes its magnitude and the unit vector $\bm{m}$ with the components
$m_x$, $m_y$, and $m_z$ its orientation. Throughout this paper, all vector components labeled
by $x$, $y$, and $z$ refer to the cubic or tetragonal coordinate system associated with the
[100], [010], and [001] crystal directions, respectively.

\subsection{Longitudinal and transverse resistivities}

We consider a standard configuration for magnetotransport measurements where the longitudinal
voltage is probed along and the transverse voltage across the direction $\bm{j}$ of a homogeneous
current with density $\bm{J}=J\bm{j}$. Accordingly, we introduce a right-handed coordinate system
with unit vectors $\bm{j}$, $\bm{t}$, and $\bm{n}$ = $\bm{j}\times\bm{t}$, so that $\bm{t}$ defines
the transverse direction and $\bm{n}$ typically the surface normal. The measured voltages arise
from the components $E_\mathrm{long} = \bm{j}\cdot\bm{E}$ and $E_\mathrm{trans} = \bm{t}\cdot\bm{E}$
of the electric field $\bm{E}$. Starting from Ohm's law $\bm{E}=\bar{\rho}\cdot \bm{J}$ with the
resistivity tensor $\bar{\rho}$, the longitudinal resistivity $\rho_{\mathrm{long}}$ (sheet
resistivity) and transverse resistivity $\rho_{\mathrm{trans}}$ (Hall resistivity) can be written
as
\begin{equation}\label{define_resistivities}
\rho_{\mathrm{long}} = \frac{E_\mathrm{long}}{J} = \bm{j} \cdot \bar{\rho} \cdot \bm{j}\;,
\quad
\rho_{\mathrm{trans}} = \frac{E_\mathrm{trans}}{J} = \bm{t} \cdot \bar{\rho} \cdot \bm{j}\;.
\end{equation}

\subsubsection{Resistivity tensor}

Following the ansatz of Birss,\cite{Bir64} the dependence of $\rho_{\mathrm{long}}$ and
$\rho_{\mathrm{trans}}$ on the magnetization orientation $\bm{m}$ is derived in a phenomenological
approach by writing the components $\rho_{ij}$ of the resistivity tensor $\bar{\rho}$ as series
expansions with respect to $m_x$, $m_y$, and $m_z$. In Ref.~\onlinecite{Lim06} we have presented
analytical expressions for $\rho_{\mathrm{long}}$ and $\rho_{\mathrm{trans}}$ including terms up
to the second order. Now, forced by the experimental results, we substantially extend this model
by taking into account terms up to the fourth order. Using the Einstein summation convention, the
series expansions read as
\begin{equation} \label{series_expansion}
 \rho_{ij} = a_{ij} + a_{kij}m_k + a_{klij}m_km_l + a_{klmij}m_km_lm_m + \ldots\;,
\end{equation}
where the components of the galvanomagnetic tensors $a_{ij}$, $a_{kij}$, ... appear as expansion
coefficients. Neumann's principle requires that a tensor representing a macroscopic physical
property of a crystal must be invariant under all symmetry operations $S$ of the corresponding
point group. The mathematical formulation of this requirement leads to conditional equations for
the expansion coefficients reading as
\begin{eqnarray} \label{neumanns_principle}
a_{ij} &=& S_{io}S_{jp}a_{op}\;, \nonumber \\
a_{kij} &=& |\bar{S}|S_{kq}S_{io}S_{jp}a_{qop}\;, \nonumber \\
a_{klij} &=& |\bar{S}|^2 S_{kq}S_{lr}S_{io}S_{jp}a_{qrop}\;, \nonumber \\
a_{klmij} &=& |\bar{S}|^3 S_{kq}S_{lr}S_{mt}S_{io}S_{jp}a_{qrtop}\;, \nonumber \\
a_{klmnij} &=& |\bar{S}|^4 S_{kq}S_{lr}S_{mt}S_{nu}S_{io}S_{jp}a_{qrtuop}\;,
\end{eqnarray}
where $S_{ij}$ denotes the $ij$ component and $|\bar{S}|$ the determinant of the symmetry matrix
$\bar{S}$. The determinant appears in the formula of Eqs.~(\ref{neumanns_principle}) since
$\bm{M}$ is an axial vector, whereas $\bm{E}$ and $\bm{J}$ are polar vectors. In order to derive
a complete set of conditional equations for the expansion coefficients, it is sufficient to apply
Eqs.~(\ref{neumanns_principle}) to a small set of generating symmetry matrices. In the cases of
cubic symmetry $T_{d}$ and tetragonal symmetry $D_{2d}$ this set consists of only two matrices,
namely $S_8$,$S_9$ and $S_2$,$S_8$, respectively. The matrices are given by\cite{Bir64}
\begin{eqnarray} \label{generating_matices}
S_2 &=&
\left( \begin{array}{ccc}
  -1 & 0 & 0 \\
  0 & 1 & 0 \\
  0 & 0 & -1 \\
\end{array} \right)\;, \quad
S_8 =
\left( \begin{array}{ccc}
  0 & -1 & 0 \\
  1 & 0 & 0 \\
  0 & 0 & -1 \\
\end{array} \right) \;, \nonumber \\
S_9 &=&
\left( \begin{array}{ccc}
  0 & 1 & 0 \\
  0 & 0 & 1 \\
  1 & 0 & 0 \\
\end{array} \right)\;.
\end{eqnarray}
Insertion of the generating matrices into Eqs.~(\ref{neumanns_principle}) reveals that most of
the expansion coefficients are equal in pairs or vanish. The resulting resistivity tensor for
tetragonal symmetry can be separated into two terms
\begin{equation} \label{rho_tetragonal}
\bar{\rho}_{\mathrm{tetragonal}} = \bar{\rho}_{\mathrm{cubic}} + \Delta\bar{\rho}\;,
\end{equation}
where $\bar{\rho}_{\mathrm{cubic}}$ is the resistivity tensor for cubic symmetry and
$\Delta\bar{\rho}$ a difference term which vanishes in the case of perfect cubic symmetry.
Written in ascending powers of $m_x$, $m_y$, and $m_z$, the two parts of
$\bar{\rho}_{\mathrm{tetragonal}}$ are given by
\begin{widetext}
\begin{eqnarray} \label{rho_cubic}
\nonumber
\bar{\rho}_{\mathrm{cubic}} &=&
A
\left( \begin{array}{ccc}
  1 & 0 & 0 \\
  0 & 1 & 0 \\
  0 & 0 & 1 \\
\end{array} \right) +
B
\left( \begin{array}{ccc}
  0 & m_z & -m_y \\
  -m_z & 0 & m_x \\
  m_y & -m_x & 0 \\
\end{array} \right) +
C_1
\left( \begin{array}{ccc}
  m_x^2 & 0 & 0 \\
  0 & m_y^2 & 0 \\
  0 & 0 & m_z^2 \\
\end{array} \right) +
C_2
\left( \begin{array}{ccc}
  0 & m_xm_y & m_xm_z \\
  m_xm_y & 0 & m_ym_z \\
  m_xm_z & m_ym_z & 0 \\
\end{array} \right) \nonumber \\
&+&
D
\left( \begin{array}{ccc}
  0 & m_z^3 & -m_y^3 \\
  -m_z^3 & 0 & m_x^3 \\
  m_y^3 & -m_x^3 & 0 \\
\end{array} \right) +
E_1
\left( \begin{array}{ccc}
  m_x^4 & 0 & 0 \\
  0 & m_y^4 & 0 \\
  0 & 0 & m_z^4 \\
\end{array} \right) +
E_2
\left( \begin{array}{ccc}
  m_y^2m_z^2 & 0 & 0 \\
  0 & m_x^2m_z^2 & 0 \\
  0 & 0 & m_x^2m_y^2 \\
\end{array} \right) \nonumber \\
&+&
E_3
\left( \begin{array}{ccc}
  0 & m_xm_ym_z^2 & m_xm_y^2m_z \\
  m_xm_ym_z^2 & 0 & m_x^2m_ym_z \\
  m_xm_y^2m_z & m_x^2m_ym_z & 0 \\
\end{array} \right)\;,
\end{eqnarray}
\begin{eqnarray} \label{delta_rho}
\Delta\bar{\rho}  &=&
\left( \begin{array}{ccc}
  0 & 0 & 0 \\
  0 & 0 & 0 \\
  0 & 0 & a \\
\end{array} \right)
 +
\left( \begin{array}{ccc}
  0 & b m_z & 0 \\
  -b m_z & 0 & 0 \\
  0 & 0 & 0 \\
\end{array} \right) +
\left( \begin{array}{ccc}
  c_3 m_z^2 & 0 & c_2 m_xm_z \\
  0 & c_3 m_z^2 & c_2 m_ym_z \\
  c_2 m_xm_z & c_2 m_ym_z & c_1 m_z^2 \\
\end{array} \right) +
\left( \begin{array}{ccc}
  0 & d_1 m_z^3 & -d_2 m_ym_z^2 \\
  -d_1 m_z^3 & 0 & d_2 m_xm_z^2 \\
  d_2 m_ym_z^2 & -d_2 m_xm_z^2 & 0 \\
\end{array} \right) \nonumber \\
&+&
\left( \begin{array}{ccc}
  e_2 m_y^2m_z^2 +e_4 m_z^4 & e_3 m_xm_ym_z^2 & e_6 m_xm_y^2m_z + e_7 m_xm_z^3 \\
  e_3 m_xm_ym_z^2 & e_2 m_x^2m_z^2 +e_4 m_z^4 & e_6 m_x^2m_ym_z + e_7 m_ym_z^3 \\
  e_6 m_xm_y^2m_z + e_7 m_xm_z^3 & e_6 m_x^2m_ym_z + e_7 m_ym_z^3 & e_5 m_x^2m_y^2 + e_1 m_z^4 \\
\end{array} \right)\;.
\end{eqnarray}
\end{widetext}
Deriving the above equations, we repeatedly made use of the trivial identity
\begin{equation} \label{identity1}
|\bm{m}|^2 = m_x^2 + m_y^2 + m_z^2 =1\,.
\end{equation}
The expansion parameters $A$, ..., $E_3$, and $a$, ..., $e_7$ are linear combinations of
the non-vanishing expansion coefficients, reading as
\begin{eqnarray} \label{res_par_cubic}
A &=& a_{11}+a_{1122}+a_{111122} \;, \nonumber \\
B &=& a_{123} + 3a_{12223} \;, \nonumber \\
C_1 &=& a_{1111}-a_{1122}+6a_{112211}-2a_{111122} \;, \nonumber \\
C_2 &=& 2a_{1212} + 4a_{111212}  \;, \nonumber \\
D &=& a_{11123}-3a_{12223}  \;, \nonumber \\
E_1 &=& a_{111111}-6a_{112211}+a_{111122}  \;, \nonumber \\
E_2 &=& 6a_{112233} - 2a_{111122}  \;, \nonumber \\
E_3 &=& 12a_{112323} - 4a_{111212}\;.
\end{eqnarray}
\begin{eqnarray} \label{res_par_tetra}
 a &=& a_{33}-a_{11}+a_{1133}-a_{1122}+a_{111133}-a_{111122}  \;, \nonumber \\
 b &=& a_{312} - a_{123} + 3a_{11312} - 3a_{12223}  \;, \nonumber \\
 c_1 &=& a_{3333}-a_{1111}+a_{1122}-a_{1133}+6a_{113333}-6a_{112211} \nonumber \\
     & & + \; 2a_{111122}-2a_{111133}  \;, \nonumber \\
 c_2 &=& 2a_{2323}-2a_{1212}+4a_{111313}-4a_{111212} \;, \nonumber \\
 c_3 &=& a_{3311}-a_{1122}+6a_{113311}-6a_{112211}  \;, \nonumber \\
 d_1 &=& 3a_{12223}-3a_{11312}+a_{33312}-a_{11123}  \;, \nonumber \\
 d_2 &=& 3a_{13323}-3a_{12223}  \;, \nonumber \\
 e_1 &=& a_{333333}-a_{111111}+6a_{112211}-6a_{113333} \nonumber \\
     & & + \; a_{111133}-a_{111122}  \;, \nonumber \\
 e_2 &=& 6a_{113322}-6a_{112233}+6a_{112211}-6a_{113311}  \;, \nonumber \\
 e_3 &=& 12a_{123312}-12a_{112323}  \;, \nonumber \\
 e_4 &=& a_{333311}-a_{111122}+6a_{112211}-6a_{113311}  \;, \nonumber \\
 e_5 &=& 2a_{111122}-2a_{111133}  \;, \nonumber \\
 e_6 &=& 4a_{111212}-4a_{111313}  \;, \nonumber \\
 e_7 &=& 4a_{133313}-4a_{111313}\;.
\end{eqnarray}
It should be noted that the present notation is partially different from that used in
Ref.~\onlinecite{Lim06}. For cubic ferromagnets with a small tetragonal distortion along [001],
as in the case of the (Ga,Mn)As layers under investigation, the expansion parameters are
expected to linearly vary with $\varepsilon_{zz}$. The parameters $A$ and $a$, just as all
other expansion parameters, then read as
\begin{equation}\label{Aa_ezz}
A = A_{rel} + \varepsilon_{zz}A'\;, \quad a = \varepsilon_{zz}a'\;.
\end{equation}
While $a$ vanishes for zero strain, $A$ becomes identical to $A_{rel}$ describing the
resistivity of the relaxed cubic crystal.

Once the resistivity tensor $\bar{\rho}$ is known, Eqs.~(\ref{define_resistivities}) allow
to calculate $\rho_{\mathrm{long}}$ and $\rho_{\mathrm{trans}}$ for any current direction
$\bm{j}$ and any orientation $\bm{t}$ of the transverse voltage probe relative to the crystal
axes. For a concise presentation of the results, it is convenient to replace the components
$m_x$, $m_y$, and $m_z$ of $\bm{m}$ referring to the cubic or tetragonal crystal axes with
$m_j$, $m_t$, and $m_n$ referring to the more experiment-related coordinate system defined
by $\bm{j}$, $\bm{t}$, and $\bm{n}$ according to
\begin{equation}\label{mi_general}
 m_i=j_i m_j + t_i m_t + n_i m_n\;, \quad (i=x,y,z).
\end{equation}
Consequently, Eq.~(\ref{identity1}) has to be rewritten as
\begin{equation} \label{identity2}
|\bm{m}|^2 = m_j^2 + m_t^2 + m_n^2 =1\;.
\end{equation}
For the rest of the paper we exclusively focus on the most common case of a current flow parallel
to the surface of a (001)-oriented sample and examine in detail the two situations where the
current direction is along [100] and [110].

\subsubsection{$\rho_{\mathrm{long}}$ and $\rho_{\mathrm{trans}}$ for current in the (001) plane}

To cover first the general case of a current flowing along an arbitrary direction within the (001)
plane, we write $\bm{j}$, $\bm{t}$, and $\bm{n}$ as
\begin{equation} \label{j_inplane}
\bm{j} = \left( \begin{array}{c}
  \cos\alpha  \\
  \sin\alpha  \\
  0 \\
\end{array} \right), \;\;
\bm{t} =
\left( \begin{array}{c}
  -\sin\alpha  \\
  \cos\alpha  \\
  0 \\
\end{array} \right), \;\;
\bm{n} =
\left( \begin{array}{c}
  0 \\
  0 \\
  1 \\
\end{array} \right),
\end{equation}
where $\alpha$ denotes the angle between the current direction and the [100] crystal axis. Using
Eqs.~(\ref{define_resistivities}), (\ref{rho_cubic}), and (\ref{delta_rho}), the resistivities
$\rho_{\mathrm{long}}$ and $\rho_{\mathrm{trans}}$ can be written as polynomials of fourth order
in the variables $m_j$, $m_t$, and $m_n$:
\begin{widetext}
\begin{eqnarray}
\rho_{\mathrm{long}} &=&
A + \left( C_1 - C_2 + \frac12 E_1 \right) \frac{1-\cos4\alpha}4 \nonumber \\
& & + \left[ C_1 - \left( C_1 - C_2 - E_1 \right) \frac{1-\cos4\alpha}2 \right] m_j^2
 + \left[ c_3 + \tilde{E}_2 - \left( C_1 - C_2 + E_1 + \tilde{E}_2 + \tilde{E}_3 \right)
\frac{1-\cos4\alpha}4 \right] m_n^2
\nonumber \\
& & + \; E_1 \cos4\alpha \; mj^4 +
\left[ - \tilde{E}_2 + e_4 + \left( \frac12 E_1 + \tilde{E}_2 + \tilde{E}_3 \right)
\frac{1-\cos4\alpha}4 \right] m_n^4
\nonumber \\
& & - \left[ \tilde{E}_2 + \left( E_1 - \tilde{E}_2 - \tilde{E}_3 \right) \frac{1-\cos4\alpha}2
 \right] m_j^2m_n^2
\nonumber \\
& & - \; \frac12 \left( C_1 - C_2 \right) \sin4\alpha \; m_jm_t - E_1 \sin4\alpha \; m_j^3m_t
+ \frac12 \left( \tilde{E}_2 + \tilde{E}_3 \right) \sin4\alpha\; m_jm_tm_n^2 \;,
\label{rho_long_gen} \\
\rho_{\mathrm{trans}} &=&
- \tilde{B} m_n
+ \left[ C_2 + \left( C_1 - C_2 + E_1 \right) \frac{1-\cos4\alpha}2 \right] m_jm_t
- \tilde{D} m_n^3 \nonumber \\
& & + \left[ \tilde{E}_3 - \left( E_1 + \tilde{E}_2 + \tilde{E}_3 \right) \frac{1-\cos4\alpha}2
\right] m_jm_tm_n^2  \nonumber \\
& & +
\frac14 \left( C_1 - C_2 + E_1 \right) \sin4\alpha \left( m_t^2 - m_j^2 \right)
 + \frac14 \left( E_1 + \tilde{E}_2 + \tilde{E}_3 \right) \sin4\alpha
\left( m_j^2 - m_t^2 \right)mn^2 \;.
\label{rho_trans_gen}
\end{eqnarray}
\end{widetext}
For simplicity, we have introduced the abbreviations
\begin{equation} \label{abbreviations}
\begin{array}{rclrcl}
 \tilde{B} & = & B + b\;, & \quad \tilde{D} & = & D + d_1 \;, \\
 \tilde{E}_2 & = & E_2 + e_2\;, & \quad \tilde{E}_3 & = & E_3 + e_3\;.
\end{array}
\end{equation}

Equations~(\ref{rho_long_gen}) and (\ref{rho_trans_gen}) drastically simplify for current
directions along [100] and [110], referring to $\alpha = 0^{\circ}$ and
$\alpha = 45^{\circ}$, respectively. In both cases the corresponding expressions for the
resistivities take the same form, namely
\begin{eqnarray}
 \rho_{\mathrm{long}} & = & \rho_0 + \rho_1 m_j^2 + \rho_2 m_n^2 + \rho_3 m_j^4 +
 \rho_4 m_n^4 \nonumber \\
 && + \;\rho_5 m_j^2m_n^2\;, \label{rho_long} \\
\rho_{\mathrm{trans}} & = & \rho_6 m_n + \rho_7 m_jm_t + \rho_8 m_n^3 \nonumber \\
 && + \; \rho_9 m_jm_tm_n^2\;.
\label{rho_trans}
\end{eqnarray}
The parameter $\rho_0$ in Eq.~(\ref{rho_long}) may be regarded as a reference for the angular
dependence of $\rho_{\mathrm{long}}$. It represents the longitudinal resistivity for parallel
alignment between $\bm{m}$ and $\bm{t}$ where $m_j = m_n =0$. The reference direction $\bm{t}$ can
be easily changed to $\bm{j}$ or $\bm{n}$ by substituting $m_j$ or $m_n$, respectively, with the
help of Eq.~(\ref{identity2}). All resistivity parameters $\rho_i$ ($i$ = 0, ..., 9) are linear
combinations of the expansion parameters $A$, $\tilde{B}$, $C_1$, ..., and depend, according to
Eqs.~(\ref{rho_long_gen}) and (\ref{rho_trans_gen}), on the angle $\alpha$, i.e., on the direction
$\bm{j}$ of the current with respect to the crystal axes.

For $\bm{j} \parallel [100]$ and $\bm{t} \parallel [010]$ the resistivity parameters $\rho_i$ are
given by
\begin{equation} \label{rho_par_100}
\begin{array}{rclrcl}
 \rho_0 &=& A\;,                   & \quad \rho_5 &=& - \tilde{E}_2, \\
 \rho_1 &=& C_1\;,                  & \quad \rho_6 &=& -\tilde{B}, \\
 \rho_2 &=& \tilde{E}_2 + c_3\;,   & \quad \rho_7 &=& C_2, \\
 \rho_3 &=& E_1\;,                 & \quad \rho_8 &=& - \tilde{D}, \\
 \rho_4 &=& - \tilde{E}_2 + e_4\;, & \quad \rho_9 &=& \tilde{E}_3\;.
\end{array}
\end{equation}
For unstrained layers with perfect cubic crystal symmetry the expansion parameters represented by
small letters vanish and we obtain the relation $\rho_4 = \rho_5 = -\rho_2$. Accordingly,
Eq.~(\ref{rho_long}) reduces to
\begin{equation}
\rho_{\mathrm{long}} =  \rho_0 + \rho_1 m_j^2 + \rho_2 m_t^2m_n^2 + \rho_3 m_j^4\;.
\label{rho_long_cubic}
\end{equation}

For $\bm{j} \parallel [110]$ and $\bm{t} \parallel [\bar{1}10]$ the resistivity parameters
$\rho_i$ are given by
\begin{eqnarray} \label{rho_par_110}
 \rho_0 &=& A + \frac12 \left[ C_1 - C_2 \right] + \frac14 E_1, \nonumber \\
 \rho_1 &=& C_2 + E_1, \nonumber \\
 \rho_2 &=& \frac12 \left[ -C_1 + C_2 + 2 c_3 - E_1 + \tilde{E}_2 - \tilde{E}_3 \right] ,
 \nonumber \\
 \rho_3 &=& -E_1 , \nonumber \\
 \rho_4 &=& \frac12 \left[ \frac12 E_1 - \tilde{E}_2 + \tilde{E}_3 + 2e_4 \right] , \nonumber \\
 \rho_5 &=& -E_1 + \tilde{E}_3, \nonumber \\
 \rho_6 &=& -\tilde{B}, \nonumber \\
 \rho_7 &=& C_1 + E_1 , \nonumber \\
 \rho_8 &=& -\tilde{D} , \nonumber \\
 \rho_9 &=& -E_1 - \tilde{E}_2 \;.
\end{eqnarray}
Here, perfect cubic crystal symmetry yields the constraints
$2\rho_2 = \rho_1 + 3\rho_3 - \rho_5 - \rho_7 - \rho_9$ and
$4\rho_4 = - 5\rho_3 + 2\rho_5 + 2\rho_9$.

The parameters $\rho_i$ ($i$ = 0, ..., 9) may be thought of as the components of ten-dimensional
vectors $\bm{\rho}$ which for $\bm{j} \parallel [110]$ and $\bm{j} \parallel [100]$ are related
to each other by the linear transformation
\begin{equation} \label{rho_trafo}
\bm{\rho}_{[110]} = \bar{T}\cdot \bm{\rho}_{[100]}\;.
\end{equation}
The matrix $T$ is given by
\begin{equation} \label{matrix_T}
\bar{T} =
\left(%
\begin{array}{cccccccccc}
  1 &  \frac12 & . &  \frac14 & . &  .       & . & -\frac12 & . &  .       \\
  . &  .       & . &  1       & . &  .       & . &  1       & . &  .       \\
  . & -\frac12 & 1 & -\frac12 & . &  \frac12 & . &  \frac12 & . & -\frac12 \\
  . &  .       & . & -1       & . &  .       & . &  .       & . &  .       \\
  . &  .       & . &  \frac14 & 1 & -\frac12 & . &  .       & . &  \frac12 \\
  . &  .       & . & -1       & . &  .       & . &  .       & . &  1       \\
  . &  .       & . & .        & . &  .       & 1 &  .       & . &  .       \\
  . &  1       & . &  1       & . &  .       & . &  .       & . &  .       \\
  . &  .       & . & .        & . &  .       & . &  .       & 1 &  .       \\
  . &  .       & . & -1       & . &  1       & . &  .       & . &  .       \\
\end{array}%
\right)
\end{equation}
and satisfies the equation $\bar{T}^{-1}=\bar{T}$.

\subsubsection{Polar plots of $\rho_{\mathrm{long}}$ and $\rho_{\mathrm{trans}}$}

In order to graphically illustrate the dependence of $\rho_{\mathrm{long}}$ and
$\rho_{\mathrm{trans}}$ on the magnetization orientation $\bm{m}$, it is instructive to set either
$m_n$, $m_j$, or $m_t$ in Eqs.~(\ref{rho_long}) and (\ref{rho_trans}) equal to zero and to display
the resulting analytical expressions for $\rho_{\mathrm{long}}$ and $\rho_{\mathrm{trans}}$ in
polar plots.

\begin{figure}[h]
\includegraphics[scale=0.45]{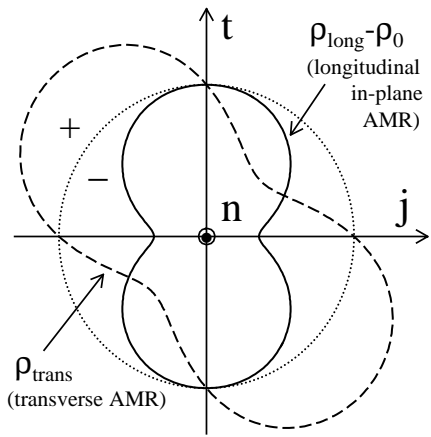}
\caption{\label{rho_polar_ip} Schematic polar plot in arbitrary units showing the dependence of
$\rho_{\mathrm{long}}-\rho_0$ (solid line) and $\rho_{\mathrm{trans}}$ (dashed line) on $\bm{m}$
for in-plane magnetization perpendicular to $\bm{n}$. The dotted circle indicates the zero line
which separates negative and positive values.}
\end{figure}

For in-plane magnetization the identity $m_n=0$ holds and Eqs.~(\ref{rho_long}) and
(\ref{rho_trans}) simplify to
\begin{eqnarray}
\rho_{\mathrm{long}} & = & \rho_0 + \rho_1 m_j^2 + \rho_3 m_j^4 \;, \label{rho_long_ip} \\
\rho_{\mathrm{trans}} & = & \rho_7 m_jm_t \label{rho_trans_ip} \;.
\end{eqnarray}
In Fig.~\ref{rho_polar_ip} the angular dependence of $\rho_{\mathrm{long}}-\rho_0$ is depicted by
the solid line, clearly reflecting the longitudinal in-plane AMR, i.e., the variation of
$\rho_{\mathrm{long}}$ with the magnetization orientation. The dashed line illustrates the angular
dependence of $\rho_{\mathrm{trans}}$ and reflects the transverse AMR. The dotted circle is the
separating line between negative and positive values of $\rho_{\mathrm{long}}-\rho_0$ and
$\rho_{\mathrm{trans}}$.

\begin{figure}[h]
\includegraphics[scale=0.45]{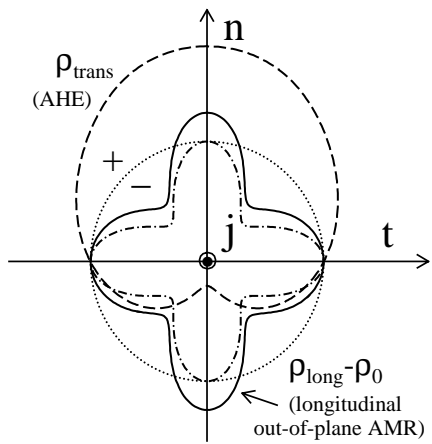}
\caption{\label{rho_polar_op_j} Dependence of $\rho_{\mathrm{long}}-\rho_0$ (solid line) and
$\rho_{\mathrm{trans}}$ (dashed line) on $\bm{m}$ for out-of-plane magnetization perpendicular
to $\bm{j}$. The dash-dotted line illustrates the special case of cubic crystal symmetry and
current along [100].}
\end{figure}

An orientation of the magnetization perpendicular to the current direction is equivalent to the
condition $m_j =0$. In this case Eqs.~(\ref{rho_long}) and (\ref{rho_trans}) reduce to
\begin{eqnarray}
\rho_{\mathrm{long}} & = & \rho_0 + \rho_2 m_n^2 + \rho_4 m_n^4 \;, \label{rho_long_op_j} \\
\rho_{\mathrm{trans}} & = & \rho_6 m_n + \rho_8 m_n^3 \label{rho_trans_op_j} \;.
\end{eqnarray}
The corresponding polar plots of $\rho_{\mathrm{long}}-\rho_0$ and $\rho_{\mathrm{trans}}$
in Fig.~\ref{rho_polar_op_j} reflect the longitudinal out-of-plane AMR for $\bm{m}\perp \bm{j}$
and the AHE, respectively. In the special case of cubic crystal symmetry and
$\bm{j} \parallel [100]$, Eq.~(\ref{rho_long_op_j}) further reduces to
\begin{equation}
\rho_{\mathrm{long}} = \rho_0 + \rho_2 m_t^2m_n^2 \;, \label{rho_long_op_sc}
\end{equation}
according to Eq.~(\ref{rho_long_cubic}) with $m_j$ = 0.
As demonstrated by the dash-dotted line in Fig.~\ref{rho_polar_op_j}, $\rho_{\mathrm{long}}$
now exhibits fourfold symmetry.

\begin{figure}[h]
\includegraphics[scale=0.45]{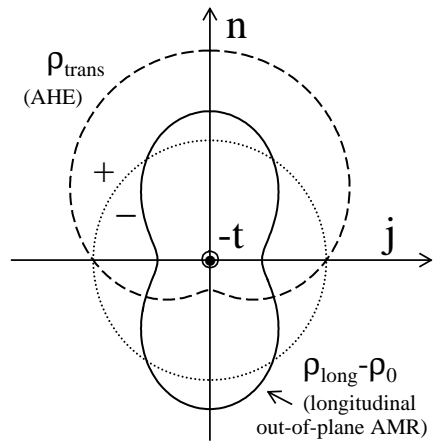}
\caption{\label{rho_polar_op_t} Schematic polar plot in arbitrary units showing the dependence of
$\rho_{\mathrm{long}}-\rho_0$ (solid line) and $\rho_{\mathrm{trans}}$ (dashed line) on $\bm{m}$
for out-of-plane magnetization perpendicular to $\bm{t}$.}
\end{figure}

The case $\bm{m}\perp \bm{t}$, or equivalently $m_t = 0$, is illustrated in
Fig.~\ref{rho_polar_op_t}, where the solid line again describes the longitudinal out-of-plane
AMR and the dashed line the AHE. The appropriate equations for $\rho_{\mathrm{long}}$ and
$\rho_{\mathrm{trans}}$ read as
\begin{eqnarray}
 \rho_{\mathrm{long}} & = & \rho_0 + \rho_1 m_j^2 + \rho_2 m_n^2 + \rho_3 m_j^4 +
 \rho_4 m_n^4 \nonumber \\
 && + \;\rho_5 m_j^2m_n^2\;, \label{rho_long_op_t} \\
\rho_{\mathrm{trans}} & = & \rho_6 m_n + \rho_8 m_n^3 \;. \label{rho_trans_op_t}
\end{eqnarray}

The expression for $\rho_{\mathrm{trans}}$ in Eqs.~(\ref{rho_trans_op_j}) and
(\ref{rho_trans_op_t}) properly describes the AHE in a phenomenological way, but it does not
include the ordinary Hall effect. For magnetic field strengths $\mu_0H < 1$\,T and hole
concentrations $p > 10^{20}$\,cm$^{-3}$ as in our experiments, however, the contribution of the
ordinary Hall effect $\mu_0H/ep$ ($e$ denotes the elementary charge) to $\rho_{\mathrm{trans}}$
is smaller than $6 \times 10^{-6}$\,$\Omega$\,cm and thus about two orders of magnitude smaller
than the measured peak values of $\rho_{\mathrm{trans}}$ (see Section IV). Effects correlated with
the magnitude $B$ of the magnetic induction $\bm{B}$, such as the negative magnetoresistance, can
be taken into account by considering $B$-dependent resistivity parameters.

\subsubsection{Polycrystalline materials}

In the literature, the longitudinal in-plane AMR and the transverse AMR in (Ga,Mn)As are often
theoretically described by the equations\cite{McG75,Jan57}
\begin{eqnarray}
\rho_{\mathrm{long}} &=& \rho_{\perp} + (\rho_{\parallel}-\rho_{\perp}) \cos^2\phi\;,
\label{amr_poly_long}\\
\rho_{\mathrm{trans}} &=& (\rho_{\parallel}-\rho_{\perp}) \sin\phi \cos\phi\;,
\label{amr_poly_trans}
\end{eqnarray}
where $\phi$ denotes the angle between $\bm{j}$ and $\bm{m}$. Taking into account the relations
$\cos^2\phi = m_j^2$ and $\sin\phi \cos\phi = m_jm_t$, which are only valid for an in-plane
configuration, it becomes clear that Eq.~(\ref{amr_poly_long}) is a good approximation to
Eq.~(\ref{rho_long_ip}) only if the fourth-order term $\rho_3 m_j^4$ is negligibly small compared
to $\rho_1 m_j^2$, i.e., if $|\rho_3| \ll |\rho_1|$, or more generally $|E_1| \ll |C_1|,|C_2|$.
The results presented in Sec. IV B reveal that these inequalities do not apply to any of the
(Ga,Mn)As samples under study. On the other hand, the angular dependence of $\rho_{\mathrm{trans}}$
described by Eq.~(\ref{amr_poly_trans}) agrees with that given by Eq.~(\ref{rho_trans_ip}). The
prefactors of the $\cos^2\phi$ and $\sin\phi \cos\phi$ terms, however, are in general different
and equal only in the special case where $\rho_1 = \rho_7$, or equivalently $C_1 = C_2$. As
already pointed out in Ref.~\onlinecite{Lim06}, Eqs.~(\ref{amr_poly_long}) and
(\ref{amr_poly_trans}) only apply to isotropic materials such as polycrystals where the
resistivities do not depend on the direction $\bm{j}$ of the current relative to the crystal
axes. In fact, Eqs.~(\ref{amr_poly_long}) and (\ref{amr_poly_trans}) result from
Eqs.~(\ref{define_resistivities}) and (\ref{rho_cubic}) by averaging over all possible spatial
orientations of the coordinate system \{$\bm{j}$,$\bm{t}$,$\bm{m}$\} with respect to the cubic
crystal axes.\cite{Bir60} Calculation yields
\begin{eqnarray} \label{rho_averaged}
\rho_{\perp} &=& A + \frac15 \left(C_1-C_2\right) + \frac1{35}\left(3E_1+3E_2-E_3\right)  \,,
\nonumber \\
\rho_{\parallel}-\rho_{\perp} &=&
\frac15 \left(2C_1+3C_2\right) + \frac1{35}\left(12E_1-2E_2+3E_3\right)\,. \nonumber \\
\end{eqnarray}

\subsubsection{Uniaxial [110] in-plane anisotropy}

Recently, Rushforth et al. presented an experimental and theoretical study on the components of the
in-plane AMR based on the equations\cite{Rus07}
\begin{eqnarray}
\frac{\rho_{\mathrm{long}}-\rho_{av}}{\rho_{av}} &=& C_I\cos2\phi + C_U\cos2\psi +
C_C\cos4\psi \nonumber \\
 &+& C_{I,C}\cos(4\psi-2\phi) \;, \label{rushforth_long} \\
\frac{\rho_{\mathrm{trans}}}{\rho_{av}} &=& C_I\sin2\phi - C_{I,C}\sin(4\psi-2\phi) \;,
\label{rushforth_trans}
\end{eqnarray}
which were obtained by extending the model of D{\"o}ring,\cite{Doe38} introduced for cubic Ni, to
systems with cubic [100] plus uniaxial [110] anisotropy. Here $\phi$ again denotes the angle
between $\bm{j}$ and $\bm{m}$, and $\psi$ the angle between $\bm{m}$ and the [110] crystal
direction. The coefficients $C_I$, $C_U$, $C_C$, and $C_{I,C}$ represent a non-crystalline,
a uniaxial, a cubic, and a crossed non-crystalline/crystalline contribution, respectively.
$\rho_{av}$ is the average value of $\rho_{\mathrm{long}}$ as $\bm{M}$ is rotated through
360$^{\circ}$. Starting from Eqs.~(\ref{rho_long_gen}) and (\ref{rho_trans_gen}) with
$m_n$ = 0, it can be shown that the coefficients $C_I$, $C_C$, $C_{I,C}$, and $\rho_{av}$
are related to the parameters $A$, $C_1$, $C_2$, and $E_1$, introduced in Sec. III A, by the
equations
\begin{eqnarray} \label{coefficients}
C_I &=& \frac1{4\rho_{av}} \left( C_1 + C_2 + E_1 \right) , \nonumber \\
C_C &=& -\frac1{8\rho_{av}} E_1 \;, \nonumber \\
C_{I,C} &=& - \frac1{4\rho_{av}} \left( C_1 - C_2 + E_1 \right) , \nonumber \\
\rho_{av} &=& A + \frac12 C_1 + \frac38 E_1  \;.
\end{eqnarray}
In our model no uniaxial [110] anisotropy of the resistivity has been taken into account, i.e.,
$C_U$ = 0, since no significant influence of such a contribution has been found in analyzing the
angle-dependent magnetotransport data. This is consistent with the fact that the strength of the
magnetic uniaxial in-plane anisotropy (for definition see Sec. III B), inferred from the
experimental data, turned out to be negligibly small in the (Ga,Mn)As samples under study. A
uniaxial in-plane contribution may become important, however, when the experiments are performed
at temperatures much higher than 4.2 K,\cite{Saw05} the thickness of the (Ga,Mn)As sample is
small,\cite{Rus07} or a uniaxial in-plain strain is induced by a piezoelectric actuator mounted
onto the sample.\cite{Goe08,Rus08} The incorporation of the contribution $C_U\cos2\psi$ from
Eq.~(\ref{rushforth_long}) into our model would yield a further term in the expression of
$\rho_{\mathrm{long}}$ in Eq.~(\ref{rho_long}). For $\bm{j} \parallel [100]$ it reads
$\rho_u m_jm_t$ and leads to an additional angular dependence of $\rho_{\mathrm{long}}$. For
$\bm{j} \parallel [110]$ it is given by $-\rho_u/2 +\rho_u m_j^2$ and can be incorporated into
the terms $\rho_0$ and $\rho_1 m_j^2$.

\subsection{Magnetic anisotropy}

In the preceeding section analytical expressions for $\rho_{\mathrm{long}}$ and
$\rho_{\mathrm{trans}}$ as a function of the magnetization orientation $\bm{m}$ have been derived.
The direction of $\bm{m}$, in turn, is determined by the MA of the ferromagnetic material and by
the strength and orientation of an external magnetic field $\bm{H}$.

Magnetic anisotropy stands for the dependence of the free energy density $F$ of a magnetic system
on the orientation of $\bm{M}$. In addition to supposing a simple single-domain model, we assume
that the magnitude $M$ of the magnetization is nearly constant under the given experimental
conditions. Instead of $F$ we therefore consider the normalized quantity $F_M$ = $F/M$, allowing
for a more concise description of the MA. For a (Ga,Mn)As film with tetragonal distortion along
[001], the anisotropic part of $F_M$ can be written as\cite{Liu06,Lim06}
\begin{eqnarray} \label{FE_a}
F_{M}(\bm{m}) & = & B_{4\parallel}\left( m_x^4 + m_y^4 \right) + B_{4\perp}m_z^4 + B_{2\perp}m_z^2
\nonumber \\
 & & + \;\frac{\mu_0M}{2}m_z^2 + B_{\bar{1}10}(m_x-m_y)^2\;.
\end{eqnarray}
The first three terms are intrinsic contributions arising from spin-orbit coupling in the valence
band. The fourth and fifth terms are extrinsic contributions describing the demagnetization
energy of an infinite plane (shape anisotropy) and a uniaxial in-plane contribution, respectively,
whose origin is still under discussion.\cite{Saw05,Wel04,Ham06} The two $m_z^2$ terms cannot be
distinguished in our experiments and are therefore lumped into a single term $B_{001}m_z^2$.

\begin{figure}[h]
\includegraphics[scale=0.6]{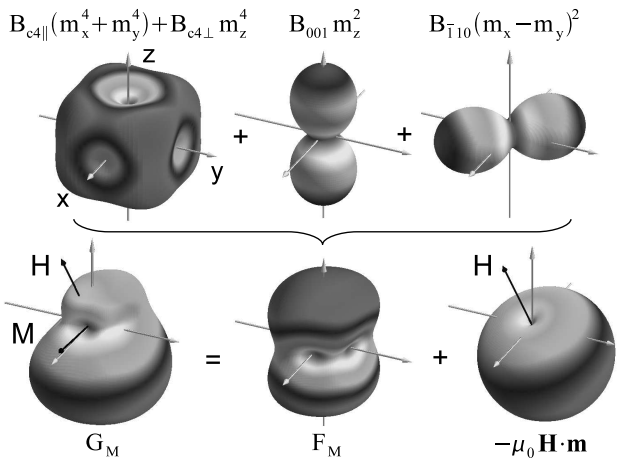}
\caption{\label{FE_sum} Graphical illustration of the different contributions to the normalized
density of the free enthalpy $G_{M}$, plotted as three-dimensional functions of the magnetization
orientation. The equilibrium position of $\bm{M}$ is determined by the minimum of $G_{M}$.}
\end{figure}

In the presence of an external magnetic field $\bm{H}$ one has to additionally take into account
the Zeeman energy, and the total energy density is finally given by the (normalized) free enthalpy
density
\begin{equation} \label{FE_b}
G_{M}(\bm{m}) = F_{M}(\bm{m}) - \mu_0\bm{H} \cdot \bm{m}\;.
\end{equation}
Figure~\ref{FE_sum} shows a graphical illustration of the various contributions to $G_{M}$. Given
an arbitrary magnitude and orientation of the magnetic field $\bm{H}$, the direction of the
magnetization $\bm{M}$ is determined by the minimum of $G_M$ with respect to the components of
$\bm{m}$.

\section{Results and discussion}

For sufficiently high magnetic fields the contribution of the free energy $F_M$ to the free
enthalpy $G_M$ in Eq.~(\ref{FE_b}) becomes much smaller than the contribution of the Zeeman
energy and can, as a good approximation, be neglected. In this case the magnetization $\bm{M}$
aligns with $\bm{H}$ and the dependence of $\rho_{\mathrm{long}}$ and $\rho_{\mathrm{trans}}$
on the magnetization orientation can be simply probed by systematically varying the direction
of $\bm{H}$. The values of the resistivity parameters $\rho_i$ are then derived by fitting
Eqs.~(\ref{rho_long}) and (\ref{rho_trans}) to the measured data with the vector components
of $\bm{m}$ replaced by those of $\bm{h}$. When the magnetic field is gradually lowered,
however, the Zeeman term in Eq.~(\ref{FE_b}) decreases, the relative contribution of $F_M$ to
$G_M$ increases, and the orientation of $\bm{M}$ more and more deviates from the direction of
$\bm{H}$ towards one of the easy axes determined by the minima of $F_M$. In other words, the
measured resistivities are increasingly influenced by the MA. In Ref.~\onlinecite{Lim06} we have
shown that this effect can be utilized to probe the MA in (Ga,Mn)As by means of angle-dependent
magnetotransport measurements.

In our experimental setup the field strength of the electromagnet is limited to 0.68\,T. In most
cases this value suffices to align $\bm{M}$ almost perfectly along $\bm{H}$. In situations,
however, where $\bm{H}$ approaches a hard magnetic axis, this might no longer apply and the
influence of the MA has to be taken into account when deriving resistivity parameters from
angle-dependent resistivity curves. Therefore, we routinely determined both resistivity and
anisotropy parameters for all samples under study. The corresponding procedure is exemplified in
Sec.~IV~A by means of an almost unstrained (Ga,Mn)As layer with $\varepsilon_{zz}$ = $-0.04\%$.
In Sec.~IV~B the strain dependence of the resistivity parameters $\rho_i$ will be discussed. An
analogous study on the anisotropy parameters $B_i$ goes beyond the scope of this paper and will
be presented elsewhere. A brief survey, however, is given in Ref.~\onlinecite{Dae08}.

\subsection{Determination of the resistivity parameters}

\begin{figure}[h]
\includegraphics[scale=0.45]{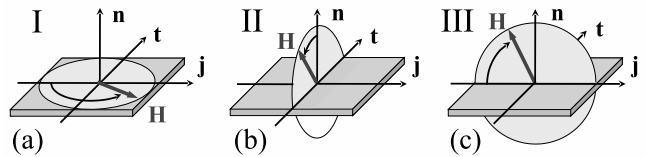}
\caption{\label{planes} The angular dependence of the resistivities was probed by rotating an
external magnetic field $\bm{H}$ within three different planes (a) perpendicular to $\bm{n}$,
(b) perpendicular to $\bm{j}$, and (c) perpendicular to $\bm{t}$. The corresponding configurations
are referred to as I, II, and III.}
\end{figure}

The longitudinal and transverse resistivities of the (Ga,Mn)As layers were measured for both
$\bm{j} \parallel [100]$ and $\bm{j} \parallel [110]$ as a function of the magnetic field
orientation at fixed field strengths of $\mu_0H$ = 0.11, 0.26, and 0.65\,T. At each field strength,
$\bm{H}$ was rotated within three different crystallographic planes perpendicular to $\bm{n}$,
$\bm{j}$, and $\bm{t}$, respectively. The corresponding configurations, labeled I, II, and III,
are shown in Fig.~\ref{planes}. In the case of $\bm{M} \parallel \bm{H}$ they are identical to
the configurations used in Figs.~\ref{rho_polar_ip}--\ref{rho_polar_op_t} to illustrate the
theoretical angular dependences of $\rho_{\mathrm{long}}$ and $\rho_{\mathrm{trans}}$ by means
of schematic polar plots.

Prior to each angular scan, the magnetization $\bm{M}$ was put into a clearly defined initial state
by raising the field to its maximum value of 0.68\,T where $\bm{M}$ is supposed to nearly saturate
and to align with the external field. The field was then lowered to one of the above mentioned
magnitudes and the scan was started.
\begin{figure*}
\includegraphics[scale=1.0]{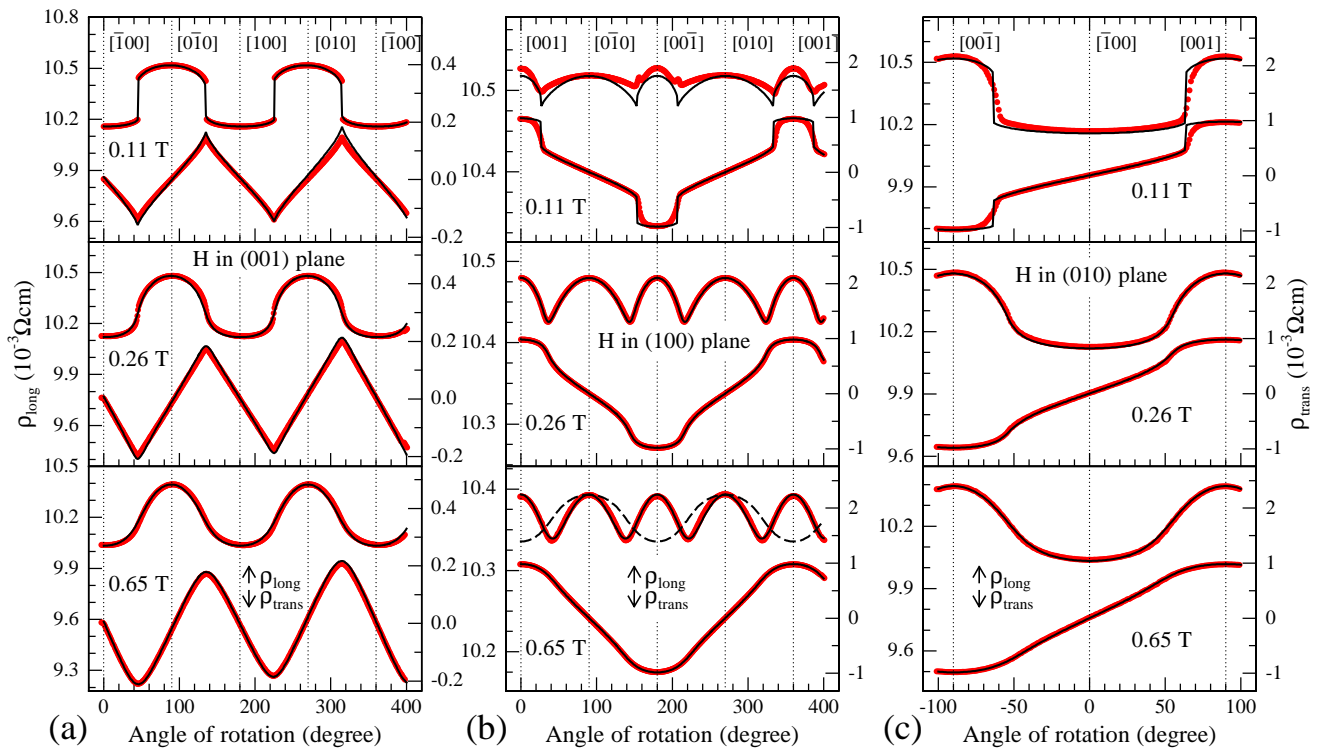}
\caption{\label{mt_B683_100} (Color online) Resistivities $\rho_{\mathrm{long}}$ and
$\rho_{\mathrm{trans}}$ recorded from the nearly unstrained (Ga,Mn)As layer with $\varepsilon_{zz}$
= $-0.04\%$ at 4.2 K and $\bm{j} \parallel [100]$ (red circles). The measurements were carried out
at fixed field strengths of $\mu_0H$ = 0.11, 0.26, and 0.65\,T with $\bm{H}$ rotated in (a) the
(001), (b) the (100), and (c) the (010) plane, corresponding to the configurations I, II, and III,
respectively. The black solid lines are fits to the experimental data using Eqs.~(\ref{rho_long})
and (\ref{rho_trans}), and one single set of resistivity and anisotropy parameters. The dashed line
in (b) at 0.65\,T represents an attempt to fit the measured curve considering only terms up to the
second order.}
\end{figure*}
Figures \ref{mt_B683_100} and \ref{mt_B683_110} show as an example the angular dependences of
$\rho_{\mathrm{long}}$ and $\rho_{\mathrm{trans}}$ for the almost unstrained sample with
$\varepsilon_{zz}$ = $-0.04\%$, measured for $\bm{j}$ along [100] and [110], respectively. The
experimental data are depicted by red circles. In the first case, $\bm{H}$ was rotated in the
(001), (100), and (010) plane, in the second case the planes of rotation were the (001), (110),
and ($\bar{1}10$) plane, corresponding to the configurations I, II, and III, respectively.
\begin{figure*}
\includegraphics[scale=1.0]{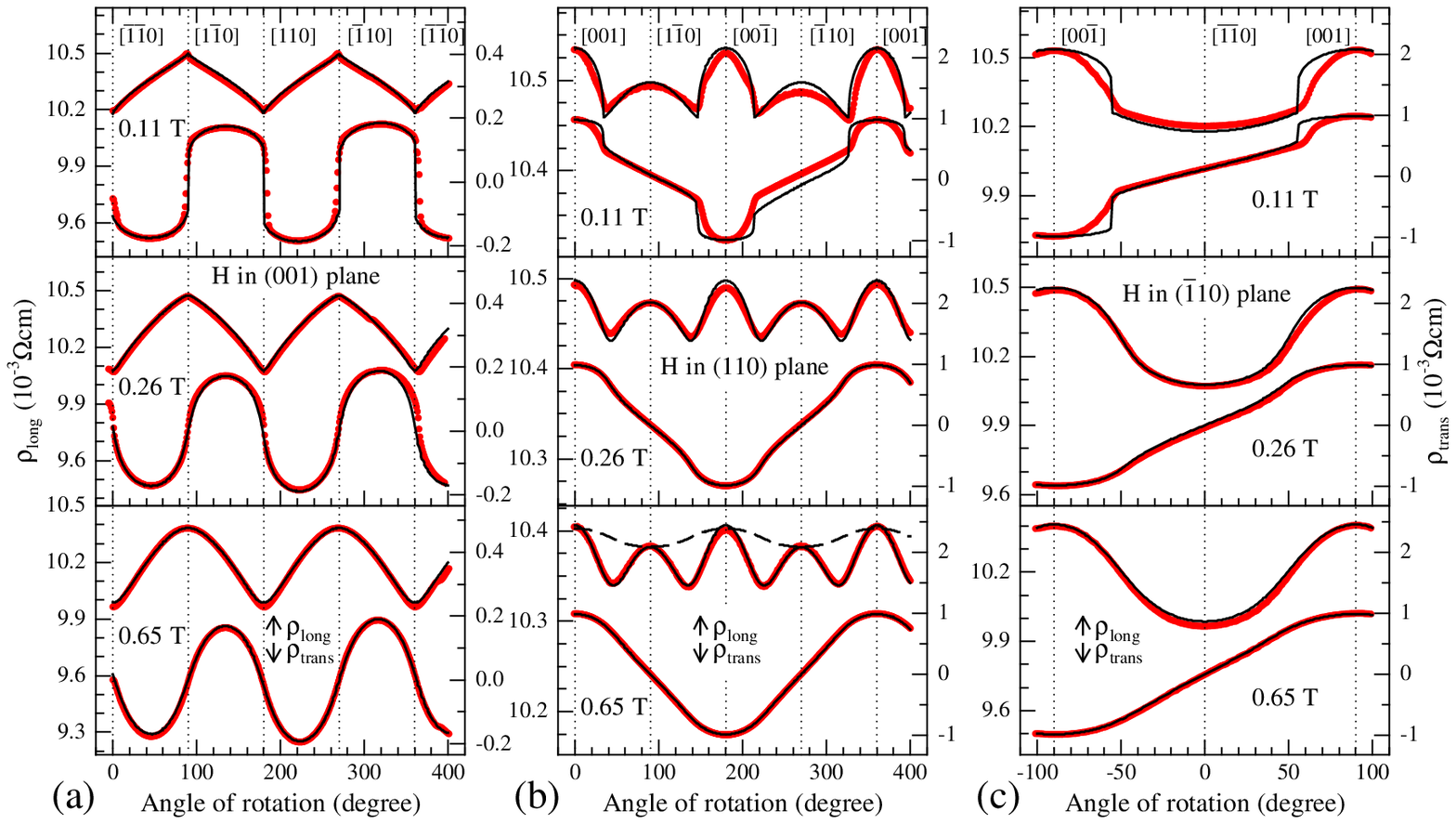}
\caption{\label{mt_B683_110} (Color online) Continuation from Fig.~\ref{mt_B683_100}. The current
direction is now along [110] and the magnetic field $\bm{H}$ was rotated in (a) the (001), (b) the
(110), and (c) the ($\bar{1}$10) plane, corresponding to the configurations I, II, and III,
respectively.}
\end{figure*}
At 0.65\,T the Zeeman energy dominates the free enthalpy. Consequently, $\bm{M}$ nearly aligns
with $\bm{H}$ and continuously follows its motion. In fact, the curves of $\rho_{\mathrm{long}}$
and $\rho_{\mathrm{trans}}$ at 0.65\,T are smooth and largely reflect the angular dependences of
the resistivities described by Eqs.~(\ref{rho_long}) and (\ref{rho_trans}) with $\bm{m}$ replaced
by $\bm{h}$. With decreasing magnetic field, the influence of the MA increases and the orientation
of $\bm{M}$ deviates more and more from the field direction. Accordingly, jumps and kinks occur in
the curves at 0.26 and 0.11\,T, arising from sudden movements of $\bm{M}$ caused by discontinuous
displacements of the minimum of $G_M$.

Values for the resistivity parameters $\rho_i$ and the anisotropy parameters $B_i$ were determined
by an iterative fit procedure. Starting with an initial guess for the anisotropy parameters, the
resistivity parameters were obtained by fitting Eqs.~(\ref{rho_long}) and (\ref{rho_trans}) to the
experimental data recorded at 0.65\,T. Then the anisotropy parameters were modified for an optimal
agreement at 0.26 and 0.11\,T, and the whole procedure was repeated until no further improvement of
the fit could be achieved. The unit vector $\bm{m}$, whose components enter Eqs.~(\ref{rho_long})
and (\ref{rho_trans}), was calculated for any given magnetic field $\bm{H}$ by numerically
minimizing $G_{M}$ with respect to the direction of $\bm{m}$.
\begin{table}
\caption{\label{tab_rho} Values of the resistivity parameters in units of $10^{-4} \Omega$~cm
determined for the nearly unstrained (Ga,Mn)As layer with $\varepsilon_{zz}$ = $-0.04\%$.}
\begin{ruledtabular}
\begin{tabular}{ccc}
$\rho_i$ ($10^{-4}$ $\Omega$\,cm) & $\bm{j} \parallel [100]$ & $\bm{j} \parallel [110]$ \\ \hline
$\rho_0$ (0.65 T)  & 103.9  & 103.8 \\
$\rho_0$ (0.26 T)  & 104.8  & 104.7 \\
$\rho_0$ (0.11 T)  & 105.2  & 105.1 \\
$\rho_1$  & -5.3  & -2.3 \\
$\rho_2$  & -2.2  & -1.9 \\
$\rho_3$  &  1.7  & -1.7 \\
$\rho_4$  &  2.2  &  2.2 \\
$\rho_5$  &  2.2  & -0.4 \\
$\rho_6$  &  9.8  &  9.8 \\
$\rho_7$  & -4.0  & -3.6 \\
$\rho_8$  &  0.0  &  0.0 \\
$\rho_9$  &  1.3  &  0.5 \\
\end{tabular}
\end{ruledtabular}
\end{table}
With the exception of $\rho_0$ the resistivity parameters were assumed to be field independent,
which turned out to be a good approximation within the accuracy of the fit. $\rho_0$ was found to
decrease with increasing magnetic field, reflecting the negative-magnetoresistance behavior of
$\rho_{\mathrm{long}}$. The values of the resistivity parameters for $\bm{j} \parallel [100]$ and
$\bm{j} \parallel [110]$, obtained from the fits, are listed in Table~\ref{tab_rho}. The two sets
of parameters are related to each other according to Eqs.~(\ref{rho_trafo}) and (\ref{matrix_T}),
in agreement with the theoretical model presented in Sec.~III. The parameter $\rho_9$ is not
immediately accessible by measurements performed in the configurations I, II, and III, since the
corresponding term in Eq.~(\ref{rho_trans}) vanishes in all three cases. It can be determined,
however, either directly by orienting $\bm{H}$ in a way that $m_j$, $m_t$, and $m_n$ are all
different from zero, or indirectly by using the relation $\rho_9' = \rho_5-\rho_3$, where the
primed and unprimed parameters correspond to $\bm{j} \parallel [110]$ and $\bm{j} \parallel [100]$,
respectively, and vice versa. For the anisotropy parameters we obtained the values $B_{4\parallel}$
= $B_{4\perp}$ = $-35$\,mT, $B_{001}$ = 35\,mT, and $B_{\bar{1}10}$ = $-5$\,mT. The theoretical
curves calculated with these parameters are drawn as solid lines in Figs.~\ref{mt_B683_100} and
\ref{mt_B683_110}.

For $\mu_0H$ = 0.26 and 0.65\,T, an excellent agreement between the measured and simulated curves
is achieved, while for $\mu_0H$ = 0.11\,T significant differences emerge. We interpret these
differences as clear evidence for the gradual breakdown of the single-domain model with constant
magnetization magnitude $M$ at low magnetic fields. The experimental curves recorded in the
configurations II and III at 0.11\,T are much smoother than the calculated ones, probably
reflecting the formation of a multitude of differently oriented ferromagnetic domains. This
interpretation is supported by a number of investigations visualizing the domain structure of
(Ga,Mn)As layers at low magnetic fields\cite{Wel03,Pro04, The06, Sug07,Dou07} as well as by
further magnetotransport measurements on the (Ga,Mn)As samples under study, not presented in this
paper. Remarkably, the theoretical curves in Figs.~\ref{mt_B683_100}(a) and \ref{mt_B683_110}(a),
calculated for configuration I where $\bm{H}$ is rotated in the (001) layer plane, almost perfectly
describe the measured curves even at 0.11\,T. This may be explained by the fact that due to shape
anisotropy the [001] axis perpendicular to the surface is a hard axis even in the unstrained layer
and that, as a consequence, ferromagnetic domains with in-plane magnetization remain more stable
at low magnetic fields than domains with out-of-plane magnetization. Accordingly, the jumps and
kinks in Figs.~\ref{mt_B683_100}(a) and \ref{mt_B683_110}(a) at 0.11\,T are extremely well
pronounced, reflecting the occurence of a nearly perfect coherent switching of the whole spin
system.

Whereas the measured resistivity curves presented in Ref.~\onlinecite{Lim06} for two (Ga,Mn)As
layers, grown on GaAs(001) and GaAs(113) substrates, could be satisfactorily well fitted by
analytical expressions containing only terms up to the second order in the components of
$\bm{m}$, it is now compulsory to take into account higher-order terms. In order to illustrate
the importance of the fourth-order terms for a correct description of the experimental traces of
$\rho_{\mathrm{long}}$, simulations are depicted as dashed lines in Figs.~\ref{mt_B683_100}(b)
and \ref{mt_B683_110}(b) where only terms up to the second order in $m_i$ were considered. No
matter which values for $\rho_2$ in Eq.~(\ref{rho_long}) were chosen, the curves completely
failed to reproduce the measured angular dependence of $\rho_{\mathrm{long}}$.

\subsection{Strain dependence of the resistivity parameters}

It is well known that a distortion of the (Ga,Mn)As crystal lattice leads to a significant change
of the MA. Microscopically, this can be explained by a strain-induced warping of the valence bands
in addition to that caused by spin-orbit coupling.\cite{Die01} The warping, however, not only
affects the MA but also the AMR and thus the resistivity parameters $\rho_i$.

In order to examine the influence of the vertical strain $\varepsilon_{zz}$ on
$\rho_{\mathrm{long}}$ and $\rho_{\mathrm{trans}}$, angle-dependent magnetotransport measurements
analogous to those presented in Figs.~\ref{mt_B683_100} and \ref{mt_B683_110} were performed on
all (Ga,Mn)As samples under study. Using the procedure described in the previous section, the
resistivity parameters $\rho_0$, ..., $\rho_8$ for $\bm{j} \parallel [100]$ and
$\bm{j} \parallel [110]$ were determined independently of each other without taking into account
their mutual relations theoretically predicted by Eqs.~(\ref{rho_trafo}) and (\ref{matrix_T}).
\begin{figure}
\includegraphics[scale=1.0]{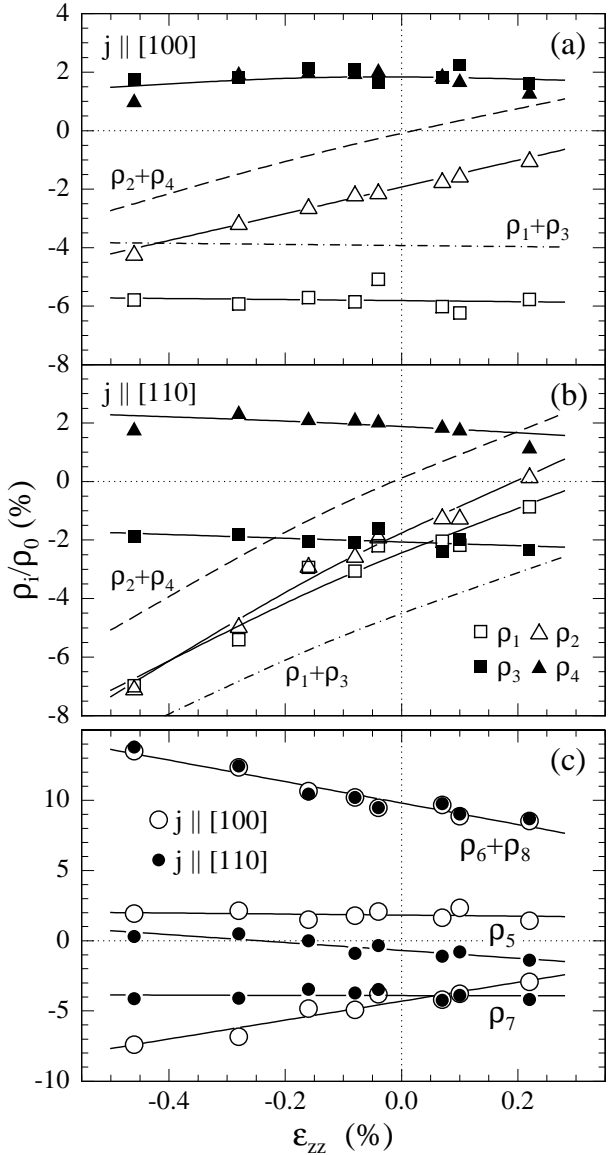}
\caption{\label{rho_ezz} Values of the normalized resistivity parameters $\rho_i/\rho_0$
($i$ = 1, ..., 8) plotted against the vertical strain $\varepsilon_{zz}$. The solid lines are
smoothing splines and are drawn to guide the eye. In (a) the normalized values of $\rho_1$, ...,
$\rho_4$ are shown for $\bm{j} \parallel [100]$ and in (b) for $\bm{j} \parallel [110]$. The
dash-dotted and dashed lines depict the strain dependence of $\rho_1+\rho_3$ and $\rho_2+\rho_4$,
respectively, and are obtained by adding the individual spline curves. In (c) the normalized
values of $\rho_5$, $\rho_6+\rho_8$, and $\rho_7$ are shown for both current directions.}
\end{figure}
\begin{figure}
\includegraphics[scale=1.0]{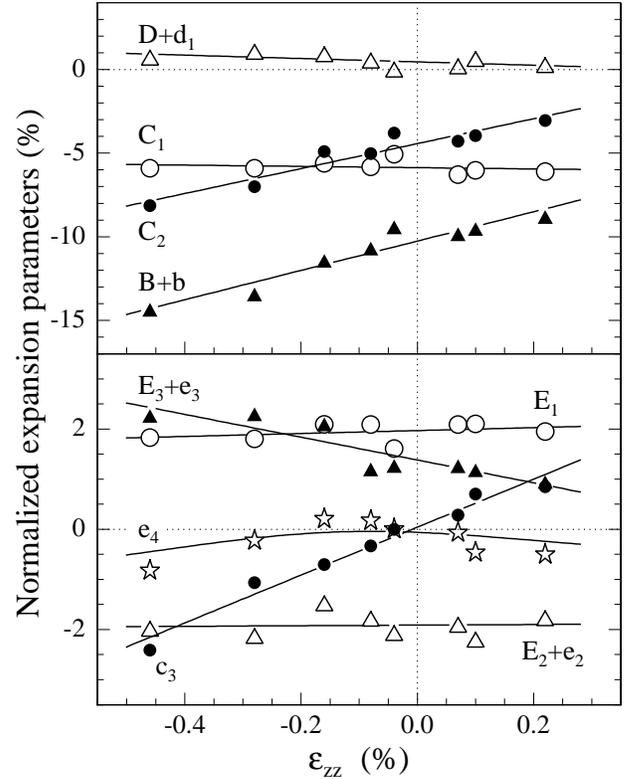}
\caption{\label{BCDE_ezz} Expansion parameters $B+b$, ..., $e_4$ plotted against the vertical
strain $\varepsilon_{zz}$. The values were calculated from $\rho_i$ ($i$ = 1, ..., 8) using
Eqs.~(\ref{rho_par_100}) and (\ref{rho_par_110}), and are normalized to that of $A$. The solid
lines are smoothing splines and are drawn to guide the eye.}
\end{figure}
Whereas the values obtained for $\rho_0$ randomly scatter in the range from
$5.0\times10^{-3}$ $\Omega$~cm to $8.8\times10^{-3}$ $\Omega$~cm, presumably due to variations
in the hole concentration and mobility, a distinct correlation with the strain is found for the
other resistivity parameters. This correlation is seen most clearly by considering the normalized
quantities $\rho_i/\rho_0$ instead of $\rho_i$. In Fig.~\ref{rho_ezz} the values of
$\rho_i/\rho_0$ ($i$ = 1, ..., 8) are plotted against $\varepsilon_{zz}$ in the range
$-0.46\% \leq \varepsilon_{zz} \leq 0.22\%$ for $\bm{j} \parallel [100]$ and
$\bm{j} \parallel [110]$. The parameter $\rho_9$ (not shown) was determined by the indirect
method described above and turned out to be nearly independent of the strain with
$\rho_9/\rho_0 \approx 0.015$ for $\bm{j} \parallel [100]$ and $\rho_9/\rho_0 \approx 0$
for $\bm{j} \parallel [110]$.

According to the theoretical model presented in Sec.~III~A, the resistivity parameters for
$\bm{j} \parallel [100]$ and $\bm{j} \parallel [110]$ should be linearly related to each other
by Eqs.~(\ref{rho_trafo}) and (\ref{matrix_T}), yielding, for instance, the symmetrical relations
\begin{equation} \label{rho_100_110}
\begin{array}{rclrcl}
  \rho'_1 & = & \rho_3 + \rho_7\;, & \quad \rho'_8 & = & \rho_8 \;,         \\
  \rho'_3 & = & -\rho_3 \;,        & \quad \rho'_7 & = & \rho_1 + \rho_3\;. \\
  \rho'_6 & = & \rho_6 \;,         &  & &
\end{array}
\end{equation}
The primed and unprimed parameters correspond to the current directions [110] and [100],
respectively, and vice versa. Inspection of Fig.~\ref{rho_ezz} reveals that the measured data
comply with the above relations supporting the validity of the theoretical model.

In contrast to the resistivity parameters $\rho_i$ the expansion parameters $\tilde{B}$, $C_1$,
..., $e_4$, do not depend on the current direction and have thus a more fundamental meaning. They
can be unambiguously calculated from the resistivity parameters using Eqs.~(\ref{rho_par_100})
and (\ref{rho_par_110}). If $\rho_9$ is measured directly, only one set of resistivity parameters
is needed, otherwise the $\rho_i$ have to be known for both current directions,
$\bm{j} \parallel [100]$ and $\bm{j} \parallel [110]$. In this work the second case applies.
Figure~\ref{BCDE_ezz} shows the values of the expansion parameters normalized to $A$ in dependence
on the vertical strain $\varepsilon_{zz}$. The smoothing spline curves suggest that the parameters
related to the cubic part $\bar{\rho}_{\mathrm{cubic}}$ of the resistivity tensor (capital letters)
as well as the parameters appearing in the strain-induced difference term $\Delta\bar{\rho}$ (small
letters) almost linearly vary with $\varepsilon_{zz}$ in agreement with Eq.~(\ref{Aa_ezz}).

\begin{figure}[h]
\includegraphics[scale=1.0]{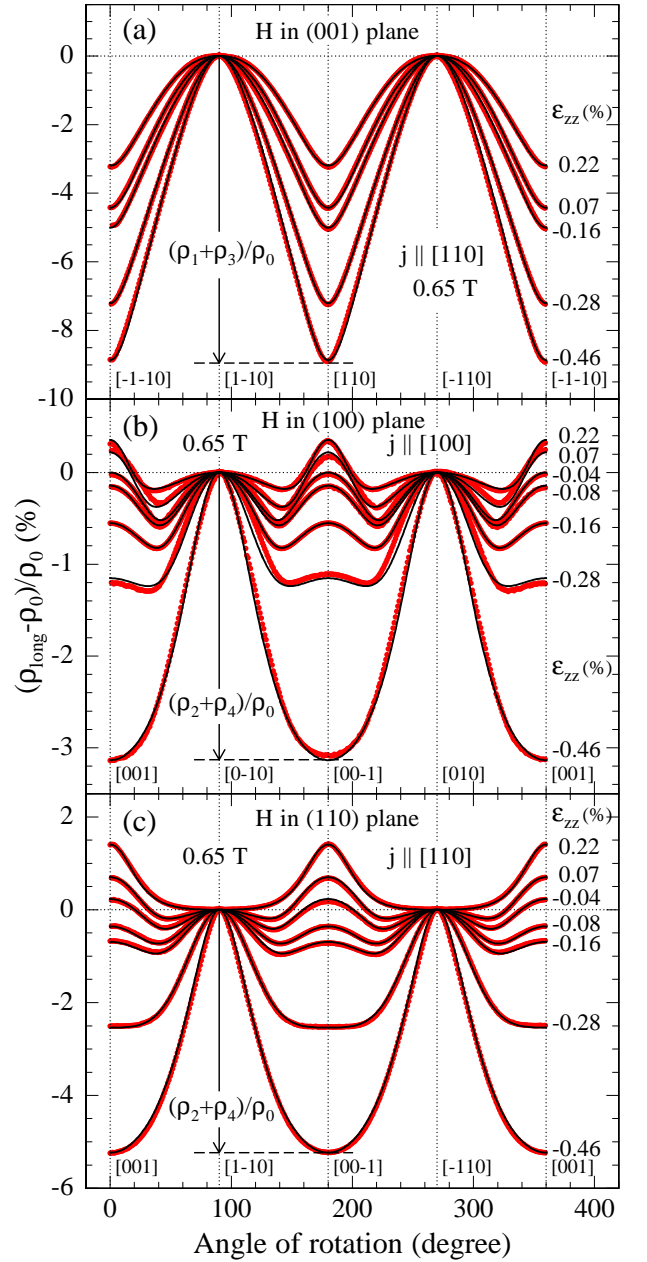}
\caption{\label{ser_0_1_long} (Color online) Variation of the angle-dependent normalized
longitudinal resistivity $(\rho_{long}-\rho_0)/\rho_0$ with vertical strain $\varepsilon_{zz}$
at $\mu_0H$ = 0.65\,T and $T$ = 4.2 K. The resistivity was recorded as a function of magnetic
field orientation for
(a) $\bm{j} \parallel [110]$ and $\bm{H}$ in the (001) plane,
(b) $\bm{j} \parallel [100]$ and $\bm{H}$ in the (100) plane, and
(c) $\bm{j} \parallel [110]$ and $\bm{H}$ in the (110) plane.
The experimental data are depicted by red circles and the calculated curves by black solid lines.}
\end{figure}

\begin{figure}
\includegraphics[scale=1.0]{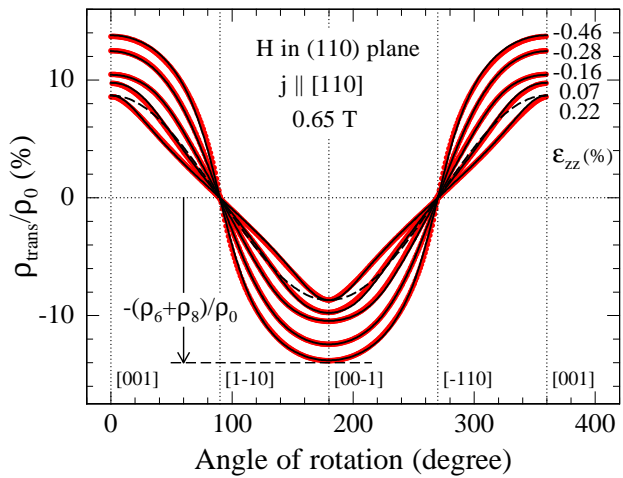}
\caption{\label{ser_1_trans} (Color online) Variation of the angle-dependent normalized transverse
resistivity $\rho_{trans}/\rho_0$ with vertical strain $\varepsilon_{zz}$ at $\mu_0H$ = 0.65\,T
and $T$ = 4.2 K. The resistivity was recorded as a function of magnetic field orientation for
$\bm{j} \parallel [110]$ and $\bm{H}$ in the (110) plane. The dashed line was calculated
setting all magnetic anisotropy parameters to zero and demonstrates the influence of the MA
at 0.65\,T for $\varepsilon_{zz}$ = 0.22\%.}
\end{figure}

The variation of the resistivity parameters with the strain manifests itself in a pronounced
change of the angular dependences of $\rho_{\mathrm{long}}$ and $\rho_{\mathrm{trans}}$. This is
exemplarily demonstrated in Figs.~\ref{ser_0_1_long} and \ref{ser_1_trans}, where the normalized
resistivities $(\rho_{long}-\rho_0)/\rho_0$ and $\rho_{trans}/\rho_0$, respectively, at
$\mu_0H$ = 0.65\,T are plotted as a function of the magnetic field orientation and the strain
for some of the configurations presented in Figs.~\ref{mt_B683_100} and \ref{mt_B683_110}.
The experimental data are again depicted by red circles and the calculated curves by black solid
lines. In Fig.~\ref{ser_0_1_long}(a) the longitudinal resistivity measured in configuration I is
shown for $\bm{j} \parallel [110]$. Provided that $\bm{M}$ and $\bm{H}$ are appoximately parallel
to each other at 0.65\,T, the angle-dependent oscillations of $\rho_{\mathrm{long}}$ reflect the
longitudinal in-plane AMR illustrated in Fig.~\ref{rho_polar_ip}. The amplitude of the oscillations
is given by
$\rho_1+\rho_3$ and increases with decreasing compressive and increasing tensile strain. It is
depicted by the dash-dotted line in Fig.~\ref{rho_ezz}(b). The specific shape of the oszillations
is determined by the ratio of $\rho_1$ and $\rho_3$ representing the $m_j^2$ and $m_j^4$ terms,
respectively. As shown in Fig.~\ref{rho_ezz}(b), $\rho_1$ strongly varies with $\varepsilon_{zz}$
whereas $\rho_3$ remains nearly constant. Figure~\ref{rho_ezz}(a) reveals that for
$\bm{j} \parallel [100]$ $\rho_1$, $\rho_3$, and $\rho_1+\rho_3$ are almost independent of the
strain. Accordingly, the longitudinal in-plane AMR is almost constant and the corresponding curves
of $\rho_{\mathrm{long}}$ are nearly identical (not shown). It should be noted again that for all
samples under study the condition $|\rho_3| \ll |\rho_1|$ does not apply and thus
Eq.~(\ref{amr_poly_long}) fails to correctly describe the angular dependence of
$\rho_{\mathrm{long}}$.

Whereas for configuration I (see above) and configuration III (not shown) the overall angular
dependence of $\rho_{\mathrm{long}}$ is basically the same in the whole range of $\varepsilon_{zz}$
under investigation, an essential variation with the strain occurs when $\bm{H}$ and thus $\bm{M}$
is rotated perpendicular to $\bm{j}$ (configuration II). In Figs.~\ref{ser_0_1_long}(b) and (c)
this variation is shown for $\bm{j} \parallel [100]$ and $\bm{j} \parallel [110]$, respectively.
The complex angular dependences emerging for $\varepsilon_{zz} > -0.46\%$, i.e., the appearance of
more than two maxima, qualitatively agree with the polar plot depicted in Fig.~\ref{rho_polar_op_j}.
They result from a competition between the second-order term $\rho_2 m_n^2$ and the fourth-order
term $\rho_4 m_n^4$, and occur whenever $\rho_2$ and $\rho_4$ are of comparable magnitude and
opposite sign, or briefly, whenever $\rho_2+\rho_4$ is close to zero. As shown by the dashed curves
in Figs.~\ref{rho_ezz}(a) and (b), this applies for $\varepsilon_{zz} > -0.46\%$ in agreement with
Figs.~\ref{ser_0_1_long}(b) and (c). For both $\bm{j} \parallel [100]$ and $\bm{j} \parallel [110]$,
$\rho_2+\rho_4$ monotonously increases with $\varepsilon_{zz}$ and changes sign at the transition
from tensile to compressive strain. In the case of $\bm{j} \parallel [100]$ symmetry demands that
$\rho_2+\rho_4$ exactly equals zero at $\varepsilon_{zz} = 0$. The corresponding polar plot is
schematically illustrated by the dash-dotted line in Fig.~\ref{rho_polar_op_j}.

The evolution of the transverse resistivity with $\varepsilon_{zz}$ is exemplarily discussed in
Fig.~\ref{ser_1_trans} where the angular dependence of $\rho_{\mathrm{trans}}/\rho_0$ is depicted
for $\bm{j} \parallel [110]$ and $\bm{m} \perp \bm{j}$ (configuration II). For magnetization
orientation along [001], i.e., perpendicular to the layer plane, $\rho_{\mathrm{trans}}$ equals
$\rho_6 + \rho_8$. As demonstrated in Fig.~\ref{rho_ezz}(c), the sum $\rho_6 + \rho_8$ is identical
for both current directions and linearly decreases with increasing $\varepsilon_{zz}$. In
Fig.~\ref{BCDE_ezz}(a) the dependences of $\rho_6 = -(B+b)$ and $\rho_8 = -(D+d_1)$ on
$\varepsilon_{zz}$ are displayed as separate curves. They reveal that the values of $\rho_6$ and
$\rho_8$ have opposite sign and differ by more than one order of magnitude. The dominant parameter
$\rho_6$ corresponds to the linear term $m_n$ in Eq.~(\ref{rho_trans}) which is usually associated
with the AHE. The increasing deviation of the curves in Fig.~\ref{ser_1_trans} from a cosinusoidal
oscillation for increasing compressive and tensile strain is primarily due to the influence of the
magnetic anisotropy even at 0.65\,T. For comparison, the resistivity curve calculated for vanishing
MA and $\varepsilon_{zz}$ = 0.22\% is drawn as a dashed line.

\section{Summary}

Based on the expansion of the resistivity tensor with respect to the direction cosines of the
magnetization up to the fourth order, we have presented a macroscopic analytical model for the
longitudinal and transverse resistivies of single-crystalline cubic and tetragonal ferromagnets.
The model applies to arbitrary magnetization orientations and current directions. It correctly
describes the results of angle-dependent magnetotransport measurements performed on a series of
(Ga,Mn)As layers with a vertical strain gradually varied from tensile to compressive. The
resistivity parameters, obtained by fitting Eqs.~(\ref{rho_long}) and (\ref{rho_trans}) to
the experimental data, were found to systematically vary with the strain.

\section*{Acknowledgements}

This work was supported by the Deutsche Forschungsgemeinschaft under contract number Li 988/4.

\end{document}